\documentclass[sigconf]{acmart}

\usepackage{graphicx}
   \graphicspath{{./Figures/}}

\usepackage{amsmath}    
\usepackage{algorithm}
\usepackage{algorithmicx}
\usepackage{multirow}
\usepackage{svg}
\usepackage{color}
\usepackage{amsmath}
\usepackage{algpseudocode}

\DeclareGraphicsExtensions{.pdf,.jpeg,.png,.fig, .emf}
   \usepackage{subfigure}
    \usepackage{epstopdf}
    \usepackage{epsfig}
    \usepackage{subfigure}
    \usepackage{epstopdf}
    \usepackage{epsfig}

\copyrightyear{2024}
\acmYear{2024}
\setcopyright{acmlicensed}\acmConference[DAC '24]{61st ACM/IEEE Design Automation Conference}{June 23--27, 2024}{San Francisco, CA, USA}
\acmBooktitle{61st ACM/IEEE Design Automation Conference (DAC '24), June 23--27, 2024, San Francisco, CA, USA}
\acmDOI{10.1145/3649329.3655988}
\acmISBN{979-8-4007-0601-1/24/06}

\begin{document}
%
\title{\vspace{-10mm} C-Nash: A Novel Ferroelectric Computing-in-Memory \\
Architecture
for Solving Mixed Strategy Nash Equilibrium }

 \author{\small{}
         Yu Qian$^1$, 
         Kai Ni$^2$, 
         Thomas K{\"a}mpfe$^3$,
         Cheng Zhuo$^{1,4*}$
         and Xunzhao Yin$^{1,4*}$ 
          \\$^1$
          Zhejiang University, Hangzhou, China;
          $^2$University of Notre Dame, Notre Dame, USA;
          $^3$Fraunhofer IPMS, Dresden, Germany
          \\$^4$Key Laboratory of CS\&AUS of Zhejiang Province, Hangzhou, China;
          $^*$Corresponding author email: \{czhuo, xzyin1\}@zju.edu.cn
}

\renewcommand{\bibfont}{\scriptsize}

\let\oldbibliography\thebibliography
\renewcommand{\thebibliography}[1]{\oldbibliography{#1}
\setlength{\itemsep}{-0.5pt}} 


\begin{abstract}
The concept of Nash equilibrium (NE), pivotal within game theory, has garnered widespread attention across numerous industries. 
However, verifying the existence of NE poses a significant computational challenge, classified as an NP-complete problem. 
Recent advancements introduced several quantum Nash solvers aimed at identifying pure strategy NE solutions (i.e., binary solutions) by integrating slack terms into the objective function, commonly referred to as slack-quadratic unconstrained binary optimization (S-QUBO).
However, 
incorporation of slack terms into the quadratic optimization 
results in changes of the objective function, which may cause incorrect solutions. 
Furthermore, these quantum solvers only identify a limited subset of pure strategy NE solutions, and fail to address
mixed strategy NE (i.e., decimal solutions), leaving many solutions undiscovered.
In this work, we propose C-Nash, a novel 
ferroelectric computing-in-memory (CiM) architecture that can efficiently handle both pure and mixed strategy NE solutions. 
The proposed architecture consists of
(i) a transformation method that converts quadratic optimization into a MAX-QUBO form without introducing additional slack variables, thereby avoiding objective function changes;
(ii) a ferroelectric FET (FeFET) based bi-crossbar structure for storing payoff matrices and accelerating the core vector-matrix-vector (VMV) multiplications of QUBO form;
(iii) A winner-takes-all (WTA) tree implementing the MAX form and a two-phase based simulated annealing (SA) logic for searching  NE solutions.
Evaluations 
show that C-Nash has up to 68.6\% increase in the success rate for identifying NE solutions, 
finding all pure and mixed NE solutions  rather than only a portion of pure NE solutions,
compared to D-Wave based quantum approaches.
Moreover, C-Nash boasts a 
reduction 
up to 157.9X/79.0X 
in time-to-solutions compared to D-Wave 2000 Q6 and D-Wave Advantage 4.1, respectively.
\end{abstract}

\maketitle
\pagestyle{empty}

%

\section{Introduction}
\label{sec:intro}

Game theory has seen remarkable success across various applications in 
industries and businesses 
\cite{carfi2011fair, roy2010survey}. 
Nash equilibrium (NE), as a fundamental concept within game theory, has garnered significant attention across multiple fields \cite{vetta2002nash, zhao2020particle}, exemplified by its application to the well-known "Prisoner's Dilemma" 
problem 
\cite{axelrod1980effective}. 
NE captures a stable solution for the decision-making problem involving multiple players. 
Each player has a finite set of actions, aiming to find an optimal strategy to choose actions to maximize their individual payoff. At the equilibrium, each player commits to a strategy that no player can unilaterally improve their payoff by changing their chosen strategy, which represents a stable state.
NE has two forms: (1) pure strategy NE 
, where players make deterministic choices, selecting a specific action; (2) mixed strategy NE 
, where 
players can choose multiple actions with respective probabilities.

Verifying the existence of NE has been proven to be NP-complete \cite{gottlob2003pure}, placing it among the most intricate computational challenges within the NP realm.
Several NE solvers has been proposed to find NE solutions.
\cite{mangasarian1964two} has proposed a method to reformulate the task of finding an NE in a game into a quadratic optimization problem.
Upon that, two quantum-based NE solvers have been introduced \cite{khan2023calculating} to solve this problem 
by furtherly transforming it into a quadratic unconstrained binary optimization (QUBO) form \cite{okrut2021calculating}.
However, this QUBO transformation involves  
the incorporation of slack terms,
which alters the original objective function,  leading to potentially wrong NE solutions.
Moreover, the quantum solvers in \cite{okrut2021calculating} can identify only a subset of pure strategy NE solutions (i.e., binary '0' or '1' values) and do not address mixed strategy NE solutions (i.e., decimal values), thereby remaining a significant portion of NE solutions undiscovered.
These challenges inspire us to develop a more versatile and efficient NE solver.

\begin{figure}[!t]
  \centering
  \includegraphics[width=1\columnwidth]{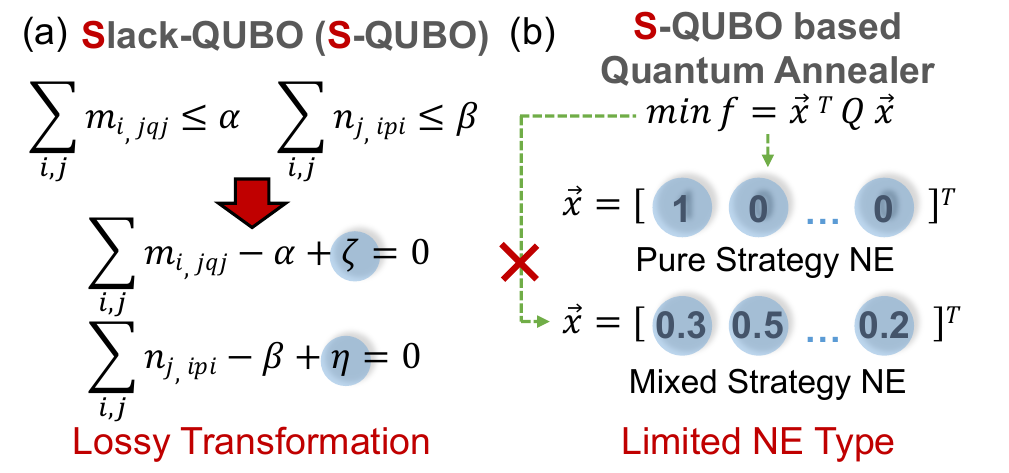}
  \vspace{-5.5ex}
  \caption{Existing approaches for addressing NE. {\bf (a)} The Slack-QUBO conversion (S-QUBO) performs a lossy transformation to handle inequalities 
  at the cost of extra slack variables; {\bf (b)} S-QUBO based quantum annealers, such as D-Wave, solve only limited type of NE.}
  \label{fig:S-QUBO}
  \vspace{-5ex}
\end{figure}

In this paper, we present C-Nash, the first ferroelectric computing-in-memory (CiM) architecture designed to handle both pure and mixed NE strategies 
with
high problem-solving efficiency.
The main contributions of this work 
can be
summarized as follows:
\begin{itemize}
    \item {
    We present a novel transformation method that converts the quadratic optimization problem \cite{mangasarian1964two} to a lossless MAX-QUBO form without introducing extra slack variables. This transformation significantly reduces the complexity of the object function (Sec. \ref{subsec:transformation}).} 
    \item {
    Leveraging the non-volatile memory property and three-terminal structure of ferroelectric FET (FeFET), we propose a FeFET-based bi-crossbar structure designed to accelerate vector-matrix-vector (VMV) multiplication, the core operation of  QUBO forms (Sec. \ref{sec:bi-crossbar}).
    } 
    \item {
    We build a winner-takes-all (WTA) tree for implementing the MAX function within the MAX-QUBO form, along with a two-phase simulated annealing (SA) logic to approach NE solutions (Sec. \ref{sec:wta} and \ref{sec:flow}).
    } 
    \item {The proposed C-Nash achieves significant improvements in both the success rate of finding NE solutions
    and the number of found solutions, 
    along with a remarkable reduction in time-to-solution compared to other Nash solvers (Sec .\ref{sec:eval}).} 
\end{itemize}


Evaluations   show that C-Nash not only showcases a remarkable 
success rate, achieving up to a 68.6\% higher rate in discovering NE solutions compared to 
D-Wave Advantage 4.1
, but also successfully identifies all NE solutions while the others only find some of pure NE solutions.
Moreover, C-Nash consumes notably minimal time costs, offering a 105.3-157.9X/18.4-79.0X less time-to-solutions compared to 
D-Wave 2000 Q6 and D-Wave Advantage 4.1, respectively.

\vspace{-1ex}
\vspace{-1ex}
\section{Background}
\label{sec:background}

In this section, we review some basics of NE, NE solvers, CiM and FeFET device.

\vspace{-1ex}
\subsection{Nash Equilibrium Basics}
\label{subsec:NE basic}

NE is a pivotal solution concept in  non-cooperative games. It represents a strategy profile in which no player has an incentive to unilaterally deviate from their respective strategic choices.
In a two-player game, an NE solution contains a pair of strategies of
player 1 and player 2
($p^\ast$, $q^\ast$) such that the expected payoff functions $f_1$ and $f_2$ for player 1 and 2, respectively, satisfy:
\begin{equation}
\vspace{-0.5ex}
\label{equ:NE}
\small f_1(p^\ast, q^\ast) \geq f_1(p, q^\ast), \ f_2(p^\ast, q^\ast) \geq f_2(p^\ast, q), \ 
\forall p,\ \forall q
\end{equation}
The payoff functions are defined as follows:
\begin{equation}
\vspace{-0.5ex}
\label{equ:payoff function}
\small f_1(p, q) = p^TMq, \ f_2(p, q) = p^TNq
\end{equation}
where $M$ and $N$ are $n \times m$ matrices representing the payoffs to player 1 and 2, respectively. $p$ and $q$ are the strategies
of player 1 and 2, respectively, i.e., $p = (p_1, p_2, ..., p_n), q = (q_1, q_2, ..., q_m)$, and $\sum_i^n p_i=1, \sum_j^m q_j=1$, where
$p_i$ and $q_j$ indicate the probability for player 1 and 2 to choose action $i$ and $j$, respectively.
A pure strategy is defined when players make their choices deterministically, i.e., both $p^\ast$ and $q^\ast$ have only one nonzero element. 
For instance, $p^\ast _1=1$ implies that player 1 chooses action $1$ as his strategy.
In contrast, a mixed strategy allows players to choose their actions with probabilities, i.e.,  $p^\ast$ and $q^\ast$  contain multiple decimal elements.
For example, the strategy $p^\ast _1=0.5, p^\ast _2=0.5$ indicates a 50\% probability for player 1 to take action $1$ and a 50\% probability for action $2$, respectively.

\begin{figure}[!t]
  \centering
  \includegraphics[width=1\columnwidth]{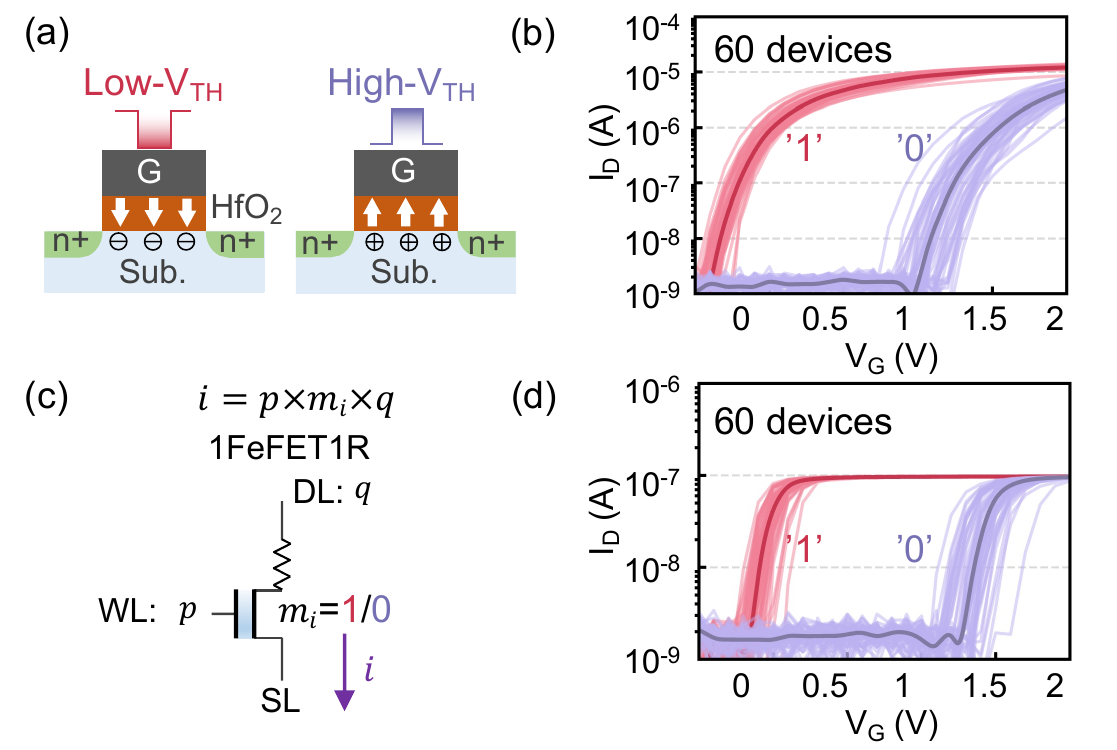}
  \vspace{-6ex}
  \caption{ 
  {\bf (a)} FeFET can be programmed to  store low/high $V_\text{TH}$ by applying different write pulses; 
  {\bf (b)} I$_D$-V$_G$ characteristics of  FeFET storing binary state $m_i$;
  {\bf (c)} The 1FeFET1R structure storing  binary $m_i$ naturally performs $i = p\times m_i \times q$ through its drain-source current $i$ by applying  two inputs $p$ and $q$ at the gate $WL$ and drain $DL$, respectively, and leveraging {\bf (d)} the  I$_D$-V$_G$ characteristics with suppressed ON current variability.
  }
  \label{fig:FeFET device}
  \vspace{-4ex}
\end{figure}

\vspace{-1ex}
\subsection{Existing NE solvers}
\label{subsec:existing work}

A method is proposed in \cite{gottlob2003pure}    to transform the state of NE (Eq. \eqref{equ:NE} and \ref{equ:payoff function}) into a quadratic optimization problem:
\begin{equation}
\vspace{-1ex}
\label{equ:qubo0}
\small \max_{p,q,\alpha,\beta} f = p^T(M+N)q - \alpha - \beta
\end{equation}
subject to
\begin{equation}
\vspace{-0.5ex}
\label{equ:qubo1}
\begin{aligned}
\small Mq-\alpha e &\leq 0\\
N^Tp - \beta l &\leq 0\\
e^T - 1 &= 0\\
l^T - 1 &= 0\\
p,q &\geq 0
\end{aligned}
\end{equation}
where $\alpha$ and $\beta$ are scalars,  $e$ and $l$ are  $n \times 1$ and $m \times 1$ vectors of ones, respectively.

To find the NE solutions of Eq. \eqref{equ:qubo0} with the constraints in Eq. \eqref{equ:qubo1}, several quantum-based NE solvers have been proposed \cite{okrut2021calculating, khan2023calculating}. 
These solvers initially transform the problem into a QUBO formula and subsequently map it onto quantum annealers to search for NE solutions.
The general QUBO form \cite{date2021qubo} is expressed as:
\begin{equation}
\vspace{-1ex}
\label{equ:QUBO}
 \small \min y = \Vec{x}\ ^TQ\Vec{x}
\vspace{-0.5ex}
\end{equation}
where $\Vec{x} \in \{0,1\}^n$, and $Q$ is an $n \times n$ matrix.
By introducing slack variables, the quadratic optimization problem  above can be transformed into QUBO formula:
\begin{equation}
\vspace{-0.5ex}
\label{equ:S-QUBO}
\begin{aligned}
 \small \min f &= - p^T(M+N)q + \alpha + \beta \\
        &+ A(\sum_i p_i - 1)^2 + B(\sum_j q_j - 1)^2 \\
        &+ C(\sum_{i,j}m_{i,j}q_j- \alpha + \zeta)^2 
        + D(\sum_{j,i}n_{i,j}p_i- \beta + \eta)^2 
\end{aligned}
\vspace{-0.5ex}
\end{equation}
where $A, B, C, D$ are constants.
We refer to this transformation as Slack-QUBO (S-QUBO). As illustrated in Fig. \ref{fig:S-QUBO}(a), S-QUBO introduces slack variables $\zeta$ and $\eta$ to equate inequalities. 
This lossy transformation changes the original constraints and  results in potential deviations in  optimal solutions.
To address this problem, we propose a novel transformation method to losslessly convert the original constraints into an equality.

The D-Wave quantum annealers are proposed for implementing S-QUBO \cite{khan2023calculating}. 
However, as shown in Fig. \ref{fig:S-QUBO}(b), the  annealer is limited to address pure strategy NE, i.e., only  two states for each variable can be represented, i.e., $x_i = 1$ or $0$, but unable to 
accommodate mixed strategy NE, which assigns decimal probability for the variables, i.e., $x_i = 0.3$.
Meanwhile, practical D-Wave annealers require expensive cryogenic cooling and exhibit limited connectivity between variables, affecting capability in addressing complex and large-scale S-QUBO \cite{albash2018demonstration, denchev2016computational, boixo2016computational}. 
To tackle these challenges, we propose to
leverage FeFET based CiM architecture to address both pure and mixed  strategy NE with improved
solving efficiency.

\vspace{-1ex}
\subsection{CiM Preliminaries and FeFET Basics}
\label{subsec:CiM Preliminary}

Crossbar structure is an essential implementation approach of CiM,
which employs non-volatile memory (NVM), e.g., resistive RAM (ReRAM) \cite{zhao2024convfifo} and magnetic tunneling junction (MTJ) \cite{mondal2018situ}, to store matrix weight elements and perform parallel vector-matrix multiplication 
operations for neural network accelerations.
However, these NVMs still possess untapped potential for enhancing energy and area efficiency. This has prompted us to explore the use of another NVM
technology
, namely FeFET, for designing the crossbar.
Compared with other NVMs, e.g., ReRAM \cite{salahuddin2018era} and MTJ \cite{zhuo2022design}, 
FeFET,  as shown in Fig. \ref{fig:FeFET device}(a),  offers several advantages, including  CMOS-compatibility, energy efficient read and write due to voltage-driven  mechanism, and compact three-terminal structure \cite{hu2021memory, huang2021computing, cai2022energy, yin2022ferroelectric, huang2023fefet, liu2022cosime, xu2023challenges}.

Fig. \ref{fig:FeFET device}(a) illustrates that FeFET can store low and high $V_\text{TH}$ states by applying negative and positive write pulses to its gate.
Fig. \ref{fig:FeFET device}(b) presents the experimentally measured I$_D$-V$_G$ characteristics of FeFET. 
Capitalizing on the characteristics, \cite{yin2023ultracompact} introduced a 1FeFET1R structure shown in Fig. \ref{fig:FeFET device}(c) to suppress the ON current variability. 
The suppressed ON current shown in Fig. \ref{fig:FeFET device}(d) enable the structure storing $m_i$ to implement consecutive multiplication, i.e., $i = p\times m_i \times q$ by applying input $p$ / $q$ at the gate $WL$ / drain $DL$. 
This operation can be extended to support VMV multiplications when 1FeFET1R cells are organized in an array, which  is exceptionally well-suited for the QUBO computations \cite{yin2024ferroelectric}. 

\begin{figure}[!t]
  \centering
  \includegraphics[width=\columnwidth]{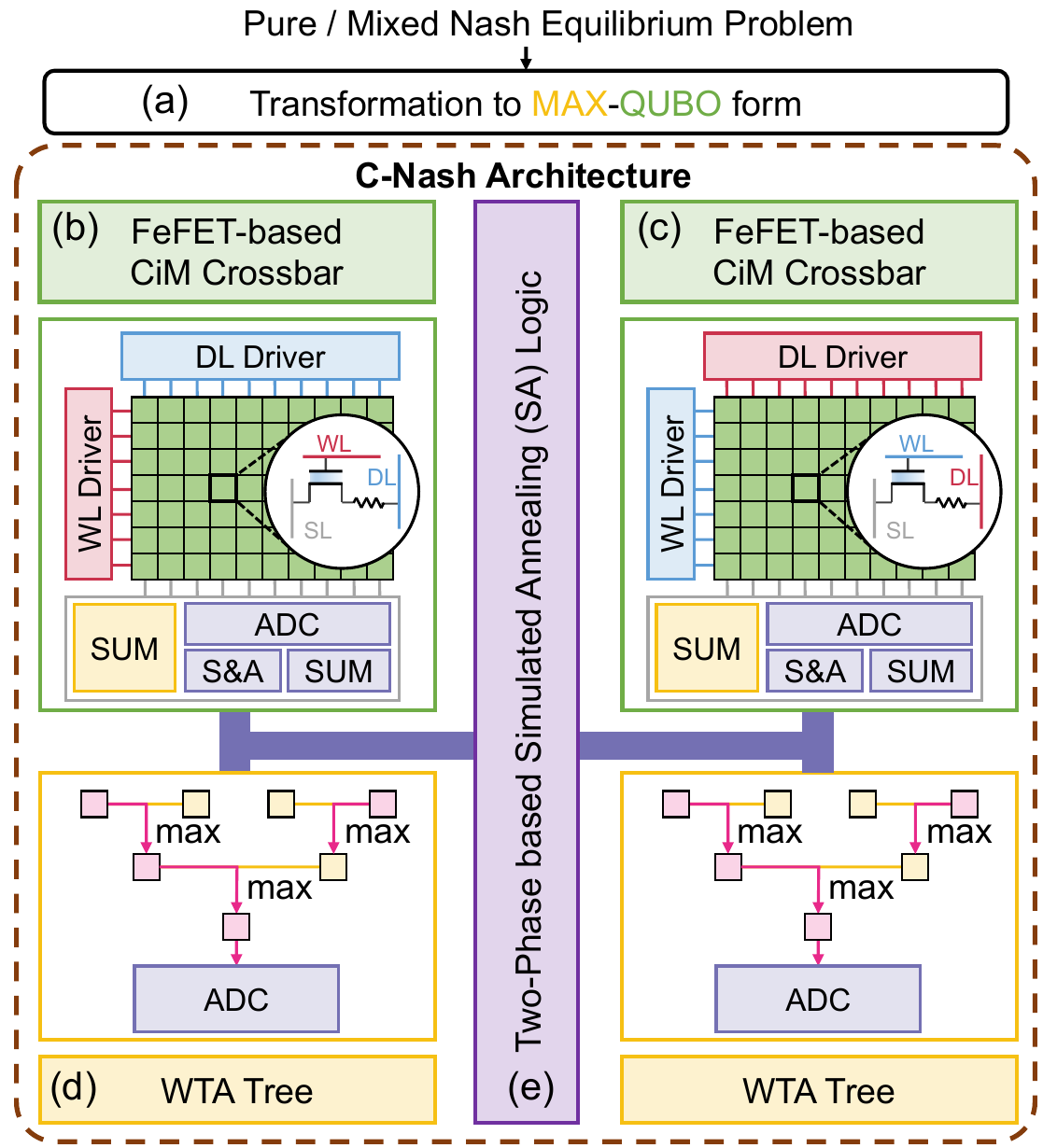}
  \vspace{-5ex}
  \caption{Overview of C-Nash. 
  (a) MAX-QUBO transformation;
  (b), (c) FeFET-based bi-crossbar structure; 
  (d) Winner-takes-all (WTA) tree;
  (e) Two-phase based simulated annealing (SA) logic.}
  \label{fig:archi overview}
  \vspace{-3ex}
\end{figure}
\vspace{-1ex}
\section{C-Nash architecture}
\label{sec:design}

In this section, we introduce C-Nash, the first ferroelectric CiM architecture  to efficiently  solve mixed strategy NE.
Fig. \ref{fig:archi overview} depicts an overview of the proposed C-Nash architecture. 
To address mixed strategy NE problems, instead of employing a lossy conversion like S-QUBO (see Sec. \ref{subsec:existing work}), C-Nash performs a lossless transformation of the quadratic optimization problem into a MAX-QUBO formulation (see Sec. \ref{subsec:transformation}), which is then mapped onto  
a FeFET-based CiM bi-crossbar structure (see Sec. \ref{sec:bi-crossbar}). A WTA tree (see Sec. \ref{sec:wta}) is built to facilitate the iterative computations of the MAX-QUBO forms, following a two-phase based SA process
(see Sec. \ref{sec:flow}).


\vspace{-1ex}
\subsection{Transformation to MAX-QUBO}
\label{subsec:transformation}


Unlike S-QUBO approach, our proposed transformation converts the two inequalities in Eq. \eqref{equ:qubo1} into two equations:
\begin{equation}
\vspace{-0.5ex}
\label{equ:alpha}
\small \alpha = \max (Mq)
\end{equation}
\begin{equation}
\vspace{-0.5ex}
\label{equ:beta}
\small \beta = \max (N^Tp) 
\end{equation}
where $Mq$ and $N^Tp$ are product vectors with $n$ and $m$ elements, respectively.
We then obtain the MAX-QUBO form as follows:
\begin{equation}
\vspace{-0.5ex}
\label{equ:max-qubo}
\small \min_{p,q} f = \max(Mq) + \max(N^Tp) - p^T(M+N)q 
\end{equation}
where $\sum_i p_i=1$ and $\sum_i q_i=1$ are satisfied by circuits.

Therefore,  given a decision-making problem (payoff matrices $M$ and $N$) and its associated strategies from two players ($p$ and $q$), the objective function can be divided into three  components: $p^T(M+N)q$, $\max(Mq)$, and $\max(N^Tp)$. Note that all three components involve matrix-vector (MV) or vector-matrix-vector (VMV) multiplications, which can be addressed perfectly by
our proposed FeFET-based CiM design.

\vspace{-1ex}
\subsection{FeFET-based Bi-Crossbar}
\label{sec:bi-crossbar}

\begin{figure}[!t]
  \centering
  \includegraphics[width=1\columnwidth]{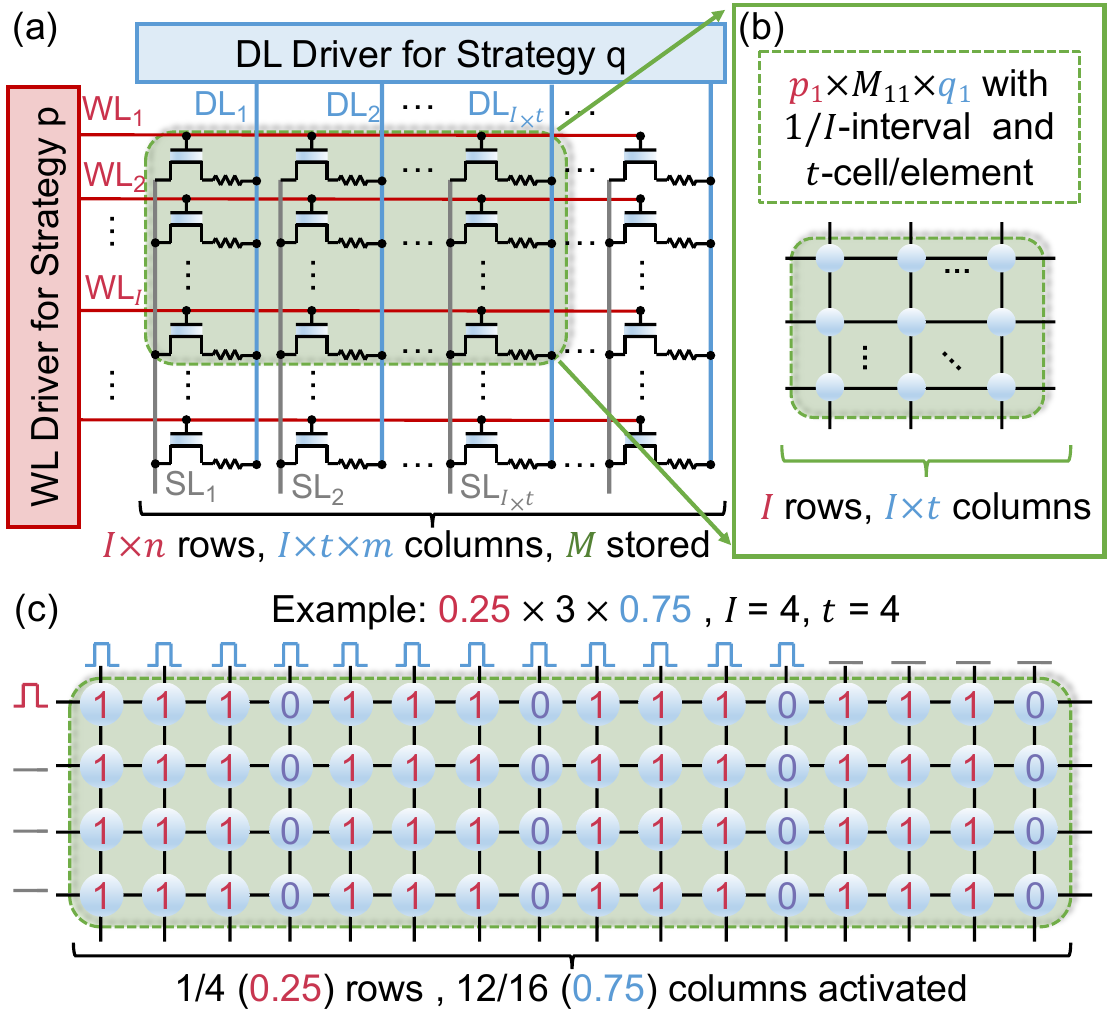}
  \vspace{-5ex}
  \caption{Mapping strategy of bi-crossbar. 
  (a) A
  FeFET based crossbar  implementing $p\times M \times q$ with strategy inputs quantified to $I$ intervals and each element of $M$ represented by $t$ cells; 
  (b) Subarray implementing scalar multiplications $p_1\times M_{11} \times q_1$;
  (c) A mapping example of $0.25\times 3 \times 0.75$ with $I=4$ and $t=4$.
  }
  \label{fig:crossbar}
  \vspace{-3ex}
\end{figure}

We employ the 1FeFET1R structure shown in Fig. \ref{fig:FeFET device} to build the CiM crossbar, as shown in Fig. \ref{fig:archi overview}(a)/(b). 
Both strategies $p$ and $q$ are applied to the inputs of crossbars, which store the payoff  matrices.
Fig. \ref{fig:crossbar}(a) depicts a comprehensive mapping of $p^TMq$ to the crossbar. 
The payoff matrix $M$ is stored within 1FeFET1R cells, and the input strategies $p$ and $q$ are encoded as driving voltages applied at the WLs and DLs,
respectively. 
The summed output currents on the SLs effectively represent the  product of $p^TMq$.
The mapping strategy assumes that
(i) each 1FeFET1R cell  stores 1 bit,  $t$ cells are used to represent a payoff matrix element, and $t$ is determined by the max value of matrix element; 
(ii) The probability of taking an action within the mixed strategies  $p$ and $q$ is quantified into $I$ intervals, indicating that $I$ rows/ $I\times t$ columns are activated to represent a decimal value in the range $\{0, 1/I, 2/I, ..., 1\}$.
Therefore, 
the size of a crossbar implementing
$p \times M \times q$ in (a) and  $p_1 \times M_{11} \times q_1$ in (b)
are  $(I\times n) \times (I\times t\times m)$ and $I \times (I\times t)$, respectively. 

Fig. \ref{fig:crossbar}(c) shows an example of (b), assuming $p_1 = 0.25$, $M_{11}=3$, $q_1 = 0.75$, i.e., a multiplication $0.25 \times 3 \times 0.75$ with $I=4$ and $t=4$. 
The required crossbar size is $4 \times 16$. 
Horizontally, four adjacent cells represent the element $3$. 
1 out of 4 total rows (i.e., 0.25) and 8 out of 12 total columns (i.e., 0.75) are activated, respectively.
The accumulated current of the crossbar  representing the product  is then measured and quantified as the MAX-QUBO form result.

\begin{figure}[!t]
  \centering
  \includegraphics[width=1\columnwidth]{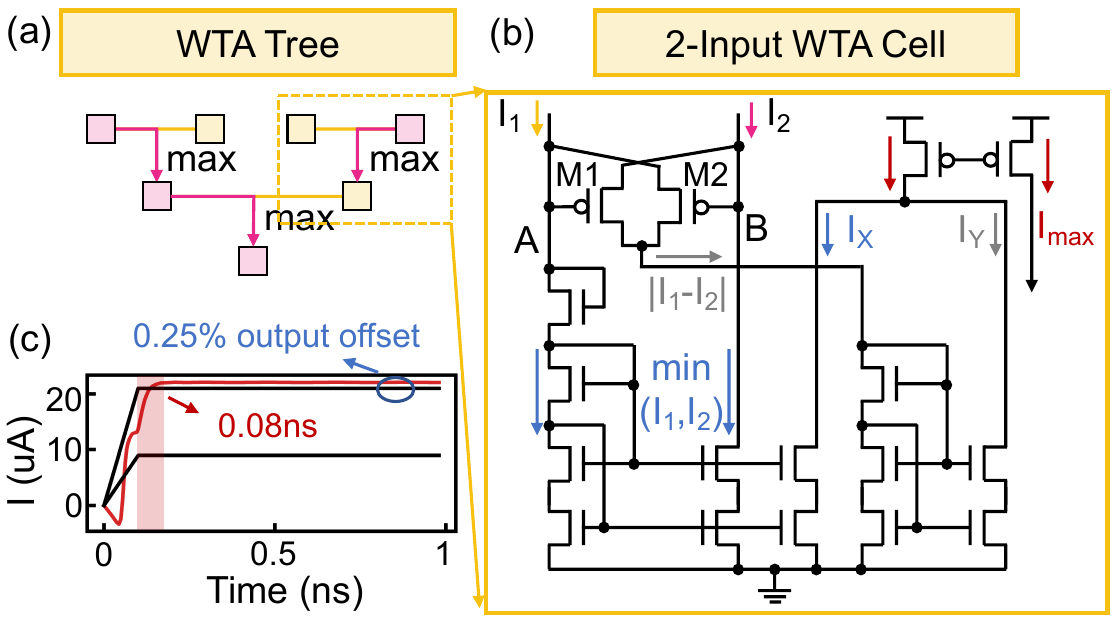}
  \vspace{-5ex}
  \caption{ 
  (a) Diagram of a WTA tree; 
  (b) Schematic of a 2-input WTA cell; 
  (c) Transient waveforms of a WTA cell.
  }
  \label{fig:wta}
  \vspace{-4ex}
\end{figure}

\vspace{-1ex}
\subsection{Winner-Takes-All Tree}
\label{sec:wta}

In C-Nash, we build a WTA tree to implement  MAX function, i.e., $\max(Mq)$, and $\max(N^Tp)$, within the MAX-QUBO form as  in Sec. \ref{subsec:transformation} for the computation of its objective function.
Fig. \ref{fig:wta}(a) shows  a WTA tree consisting of three 2-input WTA cells to obtain the maximum   among four inputs.
Generally, for $D$ inputs, i.e., $D=n$ for $\max(Mq)$, and $D=m$ for $\max(N^Tp)$,
the number  of WTA cells needed $N$ is 
$N = 2^{K-1}+2^{K-2}+...+2^0 = 2^K-1$ where $K = \lceil \log_2 D \rceil$.

Fig. \ref{fig:wta}(b) illustrates the circuit schematic of a 2-input WTA cell, which utilizes a high-swing self-biased cascode current mirror to
generate equal currents flowing through nodes A and B with  equal voltages.
When  $I_1 > I_2$, the gate voltage of M1 within the cross-coupled PMOS pair becomes higher than that of M2, turning on M2 to conduct  "extra" current $I_1 - I_2$. 
If $I_2 > I_1$, the gate voltage of M2 increases, turning on M1 to conduct "extra"
current  $I_2 - I_1$.
The smaller input current and the "extra", i.e., $\min(I_1, I_2)$ and $|I_1 - I_2|$, are then copied to 
$I_X$ and $I_Y$  through the cascode current mirror.
The  output current, $I_{\text{max}}$, is therefore given by:
\begin{equation}
\label{equ:wta}
\begin{aligned}
\small I_{\text{max}} = I_X + I_Y = \min (I_1, I_2) + |I_1 - I_2| = \max (I_1, I_2)
\end{aligned}
\end{equation}
The transient waveforms  in Fig. \ref{fig:wta}(c) 
validate the function of 2-input WTA cell, with 0.08ns latency and 
0.25\% output offset. 

\vspace{-1ex}
\subsection{Operation Flow}
\label{sec:flow}

\begin{figure}[!t]
  \centering
  \includegraphics[width=1\columnwidth]{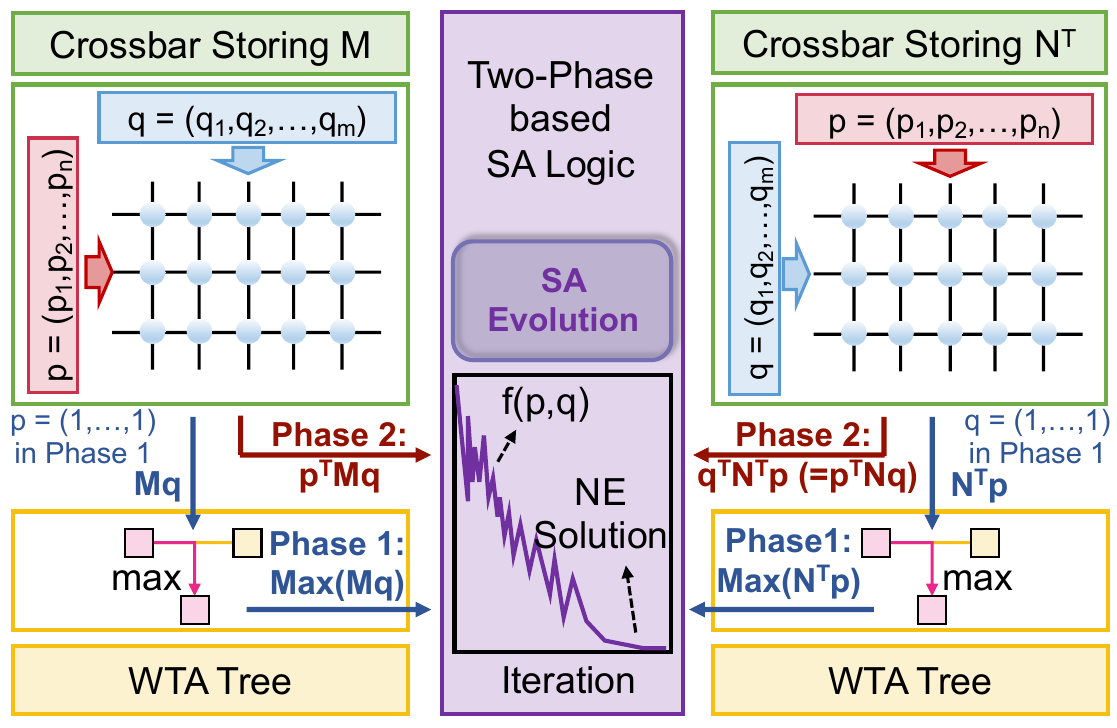}
  \vspace{-5ex}
  \caption{ Operation flow of C-Nash within an SA iteration. 
  Phase 1: Computations of $\max (Mq)$ and $\max (N^Tp)$; 
  Phase 2: Computation of $p^T(M+N)q$.
  }
  \label{fig:flow}
  \vspace{-3ex}
\end{figure}

The bi-crossbar 
and the WTA tree 
described above are used to implement one iteration of MAX-QUBO formula within a two-phase SA process.
Fig. \ref{fig:flow} depicts the operational flow. 

In Phase 1, 
both the input  vector $p$ in the  crossbar storing $M$, and $q$ in the  crossbar storing $N^T$, are configured as unit vectors, with all elements set to 1. 
Consequently, the outputs of the two crossbars are the matrix-vector multiplication  vectors, i.e., $Mq$ and $N^Tp$, respectively.
These two vectors are then fed into the current domain WTA trees to identify their maximum elements
$\max(Mq)$ and $\max(N^Tp)$, respectively.
These two maximum elements, serving as two terms in the objective function Eq. \eqref{equ:max-qubo}, are recorded
in the two-phase based SA logic for further computation.

In Phase 2, 
the bi-crossbar perform VMV multiplications, i.e., $p^TMq$ and $p^TNq$, respectively, while the WTA trees are deactivated. 
The VMV multiplication outputs are directly transmitted to the two-phase based SA logic. 
Through  addition and subtraction operations, the value of the object function  Eq. \eqref{equ:max-qubo} given a  pair of strategy $(p,q)$ is computed as an iteration result of  annealing process.

\begin{algorithm}
\caption{Simulated Annealing of C-Nash for Mixed Strategy Nash Equilibrium}
\label{alg:SA}
\begin{algorithmic}[1]
\Require 
\\$p = p_0,\ q = q_0$ \Comment{Generation of initial strategy pair}
\\$T = T_{max}$ \Comment{Starting temperature}
\Ensure Mixed strategy Nash equilibrium solution.
\State $p_c = p_0,\ q_c = q_0$
\State $f_c = f(p_c, q_c)$
\While{$T \geq T_{min}$}
\\Generate a new strategy pair $(p_{n}, q_{n})$
\\$f_{n} = f(p_{n}, q_{n})$
\\$\Delta E = f_{n} - f_c$
\If{$\Delta E  \leq 0$}
    \State $p_c = p_{n},\ q_c = q_{n}$
    \Comment{Accept the new strategy pair}
\Else
    \State Accept $(p_{n}, q_{n})$ with a probability $e^{\frac{-\Delta E}{T}}$
\EndIf
\State $T = D(T)$ \Comment{Temperature decay}
\EndWhile
\end{algorithmic}
\end{algorithm}

Alg. \ref{alg:SA} illustrates 
the SA process in C-Nash. 
During each SA iteration, the strategy pair in the last iteration randomly increment or decrement the action probabilities by the value of interval to generate  new strategy pair $(p_n, q_n)$ for MAX-QUBO computation.
The output value of the objective function given the strategy pair, denoted as $f_n$, is computed. 
The SA logic then compares $f_n$ with the objective function recorded $f_c$ and 
updates the recorded strategy pair and corresponding objective function $f_c$ per the comparison and probability related to the annealing temperature.
\vspace{-1ex}
\section{Evaluation}
\label{sec:eval}

In this section, we validate the robustness of the crossbar and WTA component of C-Nash, 
and evaluate  the problem-solving efficiency of C-Nash addressing 
three games 
in \cite{khan2023calculating}.
All simulations were performed using Cadence SPECTRE, and the Preisach FeFET model \cite{ni2018A} was adopted. 
For MOSFETs, 
the TSMC 28nm model with TT process corner was employed at 
27$^{\circ}$C. The wiring parasitics for the 28nm technology node were extracted from DESTINY \cite{poremba2015destiny}.

\vspace{-1ex}
\subsection{Robustness of C-Nash}
\label{sec:robust evaluation}

\begin{figure}[!t]
  \centering
  \includegraphics[width=1\columnwidth]{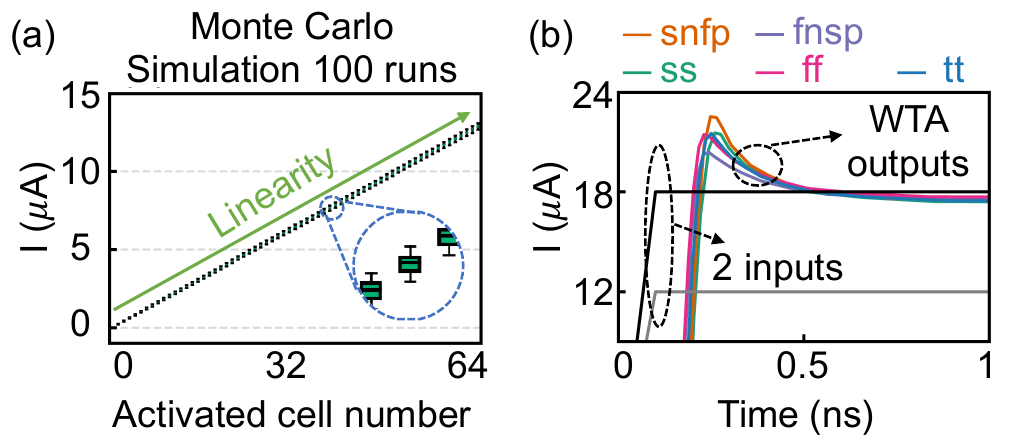}
  \vspace{-5ex}
  \caption{   
    Robustness validation on C-Nash. 
    (a) Monte Carlo simulation exhibits a good linearity of a 64$\times$64 crossbar w.r.t the number of activated cells within a column; 
    (b) Waveforms of WTA with different process corners.}
  \label{fig:mc all}
  \vspace{-4ex}
\end{figure}

We evaluate the robustness of C-Nash considering the device variability, where 
each 1FeFET1R cell of a 64$\times$ 64 crossbar assumes a device-to-device variability, i.e.,  $\sigma=40mV$ for FeFET $V_{TH}$ from \cite{soliman2023first}, and an 8\% resistor variability derived from \cite{saito2021analog}.
Fig. \ref{fig:mc all}(a) illustrates the output currents of the crossbar across 100 Monte Carlo simulations, validating a robust linearity with respect to the number of activated cells.
The output waveforms of WTA component across various process corners, including ss, snfp, fnsp, ff, and tt, shown in Fig. \ref{fig:mc all}(b), 
confirm the robustness of WTA tree in C-Nash.

\vspace{-1ex}
\subsection{ NE Solving Efficiency}
\label{sec:efficiency evaluation}

\begin{table}[!t]
\caption{Success Rates of Finding an NE Solution}
\vspace{-2ex}
\label{table:suc rate}
\centering
\resizebox{\columnwidth}{!}{
\begin{tabular}{ c | c  c  c }
\toprule
\multirow{3}*{Nash Solver} & Battle of  & \multirow{2}*{ Bird Game}  &  Modified   \\
 & the Sexes &    &  Prisoner’s Dilemma  \\
 & (2 actions) &  (3 actions)  &  (8 actions)  \\
\hline

D-Wave
& \multirow{2}*{99.62}       
& \multirow{2}*{88.16} & \multirow{2}*{-} \\
2000 Q6$^\dagger$&     &  &  \\

\hline

D-Wave
& \multirow{2}*{98.04}       
& \multirow{2}*{72.36} & \multirow{2}*{13.30} \\
Advantage 4.1$^\dagger$&     &  &  \\

\hline

C-Nash
& \multirow{2}*{100}       
& \multirow{2}*{88.94} & \multirow{2}*{81.90} \\
(this work)&     &  &  \\

\bottomrule
\end{tabular}
}
 \begin{flushleft}
 \scriptsize
$^\dagger$: Extracted from literature.

\end{flushleft}
\vspace{-3.5ex}
\end{table}

\begin{figure}[!t]
  \centering
  \includegraphics[width=1\columnwidth]{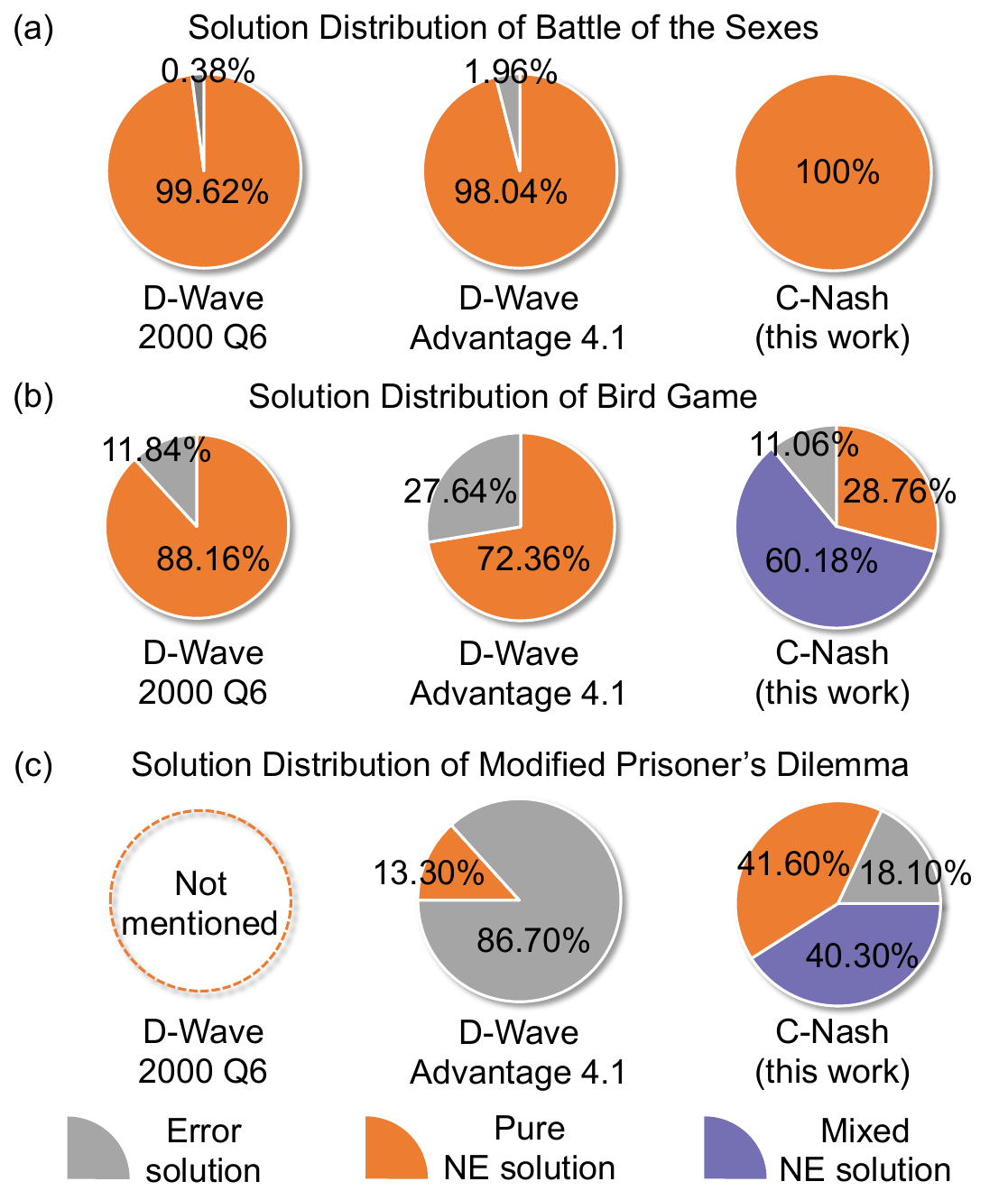}
  \vspace{-5ex}
  \caption{   
    Solution distributions of three Nash solvers in solving three games: 
    (a) Battle of the Sexes; 
    (b) Bird Game; 
    (c) Modified Prisoner's Dilemma.}
  \label{fig:solution distribution}
  \vspace{-4ex}
\end{figure}

The  problem-solving efficiencies between C-Nash and two D-Wave quantum NE solvers   \cite{khan2023calculating} are  evaluated, in terms of success rate,  number of found solutions 
(pure or mixed) and average time to find solutions. 
The benchmark dataset comprises three instances \cite{khan2023calculating}, i.e., "Battle of the Sexes", "Bird Game", and "Modified Prisoner's Dilemma", each involving two, three, and eight actions, respectively. 
The target NE solutions 
of these games 
are obtained using
Nashpy \cite{nashpyproject} as the ground truth.
The instances are executed 5000 SA runs using C-Nash, with each run comprising 10000, 15000 and 50000 iterations, respectively. 
Time required for finding solutions is derived based on 
the operational frequency  of FeFET crossbar arrays detailed in \cite{soliman2023first}, scaling to a precision of 1-bit/1-bit. 

Table \ref{table:suc rate} summarizes the success rate of finding an NE solution
using  three Nash solvers.
The two D-Wave Nash solvers exhibit success rates exceeding 90\% in addressing the relatively simple "Battle of the Sexes" game with two actions. 
Nevertheless, as the problem complexity increases, such as in the "Bird Game" with three actions, their success rates drop to 88.16\% and 72.36\%, respectively.
The D-Wave Advantage 4.1 achieves a low success rate of 13.3\% in  more complex "Modified Prisoner's Dilemma",  featuring eight actions. 
These results show that 
with increased problem complexity,  conventional transformation method S-QUBO introduces more slack variables, thus leading to larger deviations in the object function, and  degraded performance of Nash solvers.
On the contrary, C-Nash  achieves a remarkable 100\% success rate in addressing the "Battle of the Sexes", and  maintains  higher success rates, i.e., 88.94\% and 81.90\% 
in the other two games. 
This is because that the proposed lossless MAX-QUBO conversion of C-Nash preserves the integrity of  object function, eliminating the potential solution deviations.  

\begin{figure}[!t]
  \centering
  \includegraphics[width=1\columnwidth]{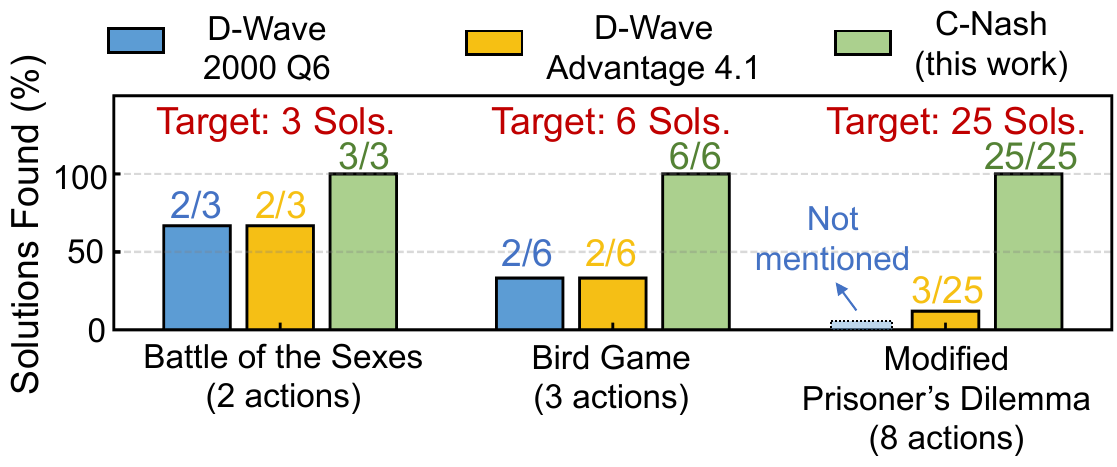}
  \vspace{-6ex}
  \caption{   
    Proportions of distinct NE solutions found by different NE solvers.
    }
  \label{fig:detail}
  \vspace{-3ex}
\end{figure}

We examine the quality of found solutions, specifically 
(i) the distributions of NE solutions identified by each NE solver across all SA runs, and (ii) the number of distinct solutions found by solvers. 
Fig. \ref{fig:solution distribution}(a)-(c) compare the distribution of  solutions found by the NE solvers in "Battle of the Sexes", "Bird Game" and "Modified Prisoner's Dilemma",  respectively. 
It can be seen that
C-Nash not only identifies pure NE solutions in all games, 
but also finds mixed NE solutions the other two solvers cannot find. 
Note that solvers may find the same solutions in different SA runs.
Fig. \ref{fig:detail} 
reveals that as the problem complexity increases,  
the proportions of distinct solutions found by the two D-Wave solvers over target (ground truth) solutions gradually  diminish, while
C-Nash consistently discovers all possible solutions.
This is because that C-Nash leverages lossless MAX-QUBO transformation to maintain the objective function's integrity, and can accommodate MAX-QUBO form with decimal values by exploiting the FeFET to support VMV multiplications.

\begin{figure}[!t]
  \centering
  \includegraphics[width=1\columnwidth]{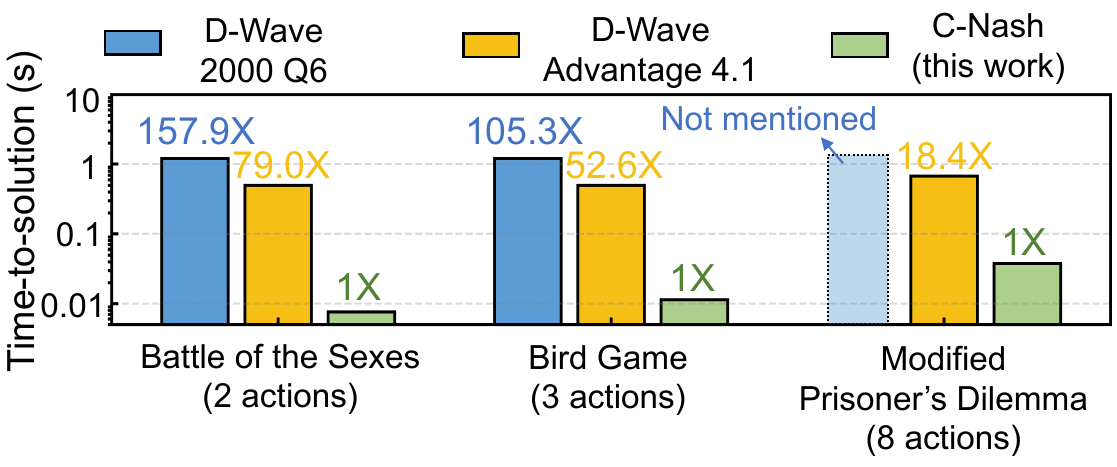}
  \vspace{-6ex}
  \caption{   
    Time to solutions of three Nash solvers.
    }
  \label{fig:time}
  \vspace{-4ex}
\end{figure}

Fig. \ref{fig:time} illustrates the average time for finding the solutions of the three Nash solvers.
Compared to the other two D-Wave-based solvers, 
C-Nash consumes much less time for finding the solutions. 
This is due to two main factors: 
(i) The S-QUBO transformation of the two D-Wave based solvers deviates the objective function, resulting in potential "fake" optimal solution that minimizes the deviated objective function but is not the NE solution for original objective function.
The integrity of the objective function in C-Nash prevents the "fake" optimal solution, thus requiring fewer SA runs to find all NE solutions. 
(ii) The core operations within an SA iteration, i.e., MV multiplication, are accelerated by the FeFET-based CiM bi-crossbar structure, resulting in significant time savings.
Therefore, C-Nash achieves a better time-to-solution.

\vspace{-1ex}
\vspace{-1ex}
\section{Conclusion}
\label{sec:conclusion}
\vspace{-1ex}

In this paper, we present C-Nash, the first ferroelectric CiM architecture designed for solving mixed strategy NE solutions. 
Our work proposes a novel transformation method that  
converts quadratic optimization problem into a MAX-QUBO form without changes in original objective function.
Leveraging the unique characteristics of FeFET, we introduce a FeFET-based bi-crossbar that accelerates the 
VMV multiplication for MAX-QUBO form, and a WTA tree that implements the MAX function.
The iterative MAX-QUBO form computations are controlled by a SA logic. 
Evaluations show  significant improvements of C-Nash than other NE solvers, including higher success rates in finding NE solutions, increased number of solutions that can be found, and less time costs for finding solutions.


\vspace{-1ex}
\section*{Acknowledgment}

\vspace{-1ex}
This work was partially supported by  National Key R\&D Program
of China (2022YFB4400300), NSFC (62104213, 92164203) and SGC Cooperation Project (Grant No. M-0612).
\vspace{-1.5ex}



%

\bibliographystyle{IEEEtran}
\bibliography{ref}

\begin{thebibliography}{10}
\providecommand{\url}[1]{#1}
\csname url@samestyle\endcsname
\providecommand{\newblock}{\relax}
\providecommand{\bibinfo}[2]{#2}
\providecommand{\BIBentrySTDinterwordspacing}{\spaceskip=0pt\relax}
\providecommand{\BIBentryALTinterwordstretchfactor}{4}
\providecommand{\BIBentryALTinterwordspacing}{\spaceskip=\fontdimen2\font plus
\BIBentryALTinterwordstretchfactor\fontdimen3\font minus \fontdimen4\font\relax}
\providecommand{\BIBforeignlanguage}[2]{{%
\expandafter\ifx\csname l@#1\endcsname\relax
\typeout{** WARNING: IEEEtran.bst: No hyphenation pattern has been}%
\typeout{** loaded for the language `#1'. Using the pattern for}%
\typeout{** the default language instead.}%
\else
\language=\csname l@#1\endcsname
\fi
#2}}
\providecommand{\BIBdecl}{\relax}
\BIBdecl

\bibitem{carfi2011fair}
D.~Carf{\`\i} \emph{et~al.}, ``Fair redistribution in financial markets: a game theory complete analysis,'' \emph{Journal of Advanced Studies in Finance}, vol.~2, no.~2, p.~4, 2011.

\bibitem{roy2010survey}
S.~Roy \emph{et~al.}, ``A survey of game theory as applied to network security,'' in \emph{2010 43rd Hawaii International Conference on System Sciences}.\hskip 1em plus 0.5em minus 0.4em\relax IEEE, 2010, pp. 1--10.

\bibitem{vetta2002nash}
A.~Vetta, ``Nash equilibria in competitive societies, with applications to facility location, traffic routing and auctions,'' in \emph{The 43rd Annual FOCS}.\hskip 1em plus 0.5em minus 0.4em\relax IEEE, 2002, pp. 416--425.

\bibitem{zhao2020particle}
C.~Zhao \emph{et~al.}, ``Particle swarm optimization algorithm with self-organizing mapping for nash equilibrium strategy in application of multiobjective optimization,'' \emph{IEEE TNNLS}, vol.~32, pp. 5179--5193, 2020.

\bibitem{axelrod1980effective}
R.~Axelrod, ``Effective choice in the prisoner's dilemma,'' \emph{Journal of conflict resolution}, vol.~24, pp. 3--25, 1980.

\bibitem{gottlob2003pure}
G.~Gottlob \emph{et~al.}, ``Pure nash equilibria: Hard and easy games,'' in \emph{Proceedings of the 9th Conference on Theoretical Aspects of Rationality and Knowledge}, 2003, pp. 215--230.

\bibitem{mangasarian1964two}
O.~L. Mangasarian \emph{et~al.}, ``Two-person nonzero-sum games and quadratic programming,'' \emph{Journal of Mathematical Analysis and applications}, vol.~9, pp. 348--355, 1964.

\bibitem{khan2023calculating}
F.~S. Khan \emph{et~al.}, ``Calculating nash equilibrium on quantum annealers,'' 2023.

\bibitem{okrut2021calculating}
O.~Okrut \emph{et~al.}, ``Calculating nash equilibrium and nash bargaining solution on quantum annealers,'' \emph{arXiv e-prints}, pp. arXiv--2112, 2021.

\bibitem{date2021qubo}
P.~Date \emph{et~al.}, ``Qubo formulations for training machine learning models,'' \emph{Scientific reports}, vol.~11, p. 10029, 2021.

\bibitem{albash2018demonstration}
T.~Albash \emph{et~al.}, ``Demonstration of a scaling advantage for a quantum annealer over simulated annealing,'' \emph{Physical Review X}, vol.~8, p. 031016, 2018.

\bibitem{denchev2016computational}
V.~S. Denchev \emph{et~al.}, ``What is the computational value of finite-range tunneling?'' \emph{Physical Review X}, vol.~6, p. 031015, 2016.

\bibitem{boixo2016computational}
S.~Boixo \emph{et~al.}, ``Computational multiqubit tunnelling in programmable quantum annealers,'' \emph{Nature communications}, vol.~7, p. 10327, 2016.

\bibitem{zhao2024convfifo}
L.~Zhao \emph{et~al.}, ``Convfifo: A crossbar memory pim architecture for convnets featuring first-in-first-out dataflow,'' in \emph{29th ASP-DAC}.\hskip 1em plus 0.5em minus 0.4em\relax IEEE, 2024, pp. 824--829.

\bibitem{mondal2018situ}
A.~Mondal \emph{et~al.}, ``In-situ stochastic training of mtj crossbar based neural networks,'' in \emph{Proceedings of the International Symposium on Low Power Electronics and Design}, 2018, pp. 1--6.

\bibitem{salahuddin2018era}
S.~Salahuddin \emph{et~al.}, ``The era of hyper-scaling in electronics,'' \emph{Nat. Electron.}, vol.~1, pp. 442--450, 2018.

\bibitem{zhuo2022design}
C.~Zhuo \emph{et~al.}, ``Design of ultra-compact content addressable memory exploiting 1t-1mtj cell,'' \emph{IEEE Transactions on Computer-Aided Design of Integrated Circuits and Systems}, 2022.

\bibitem{hu2021memory}
X.~S. Hu \emph{et~al.}, ``In-memory computing with associative memories: a cross-layer perspective,'' in \emph{2021 IEEE IEDM}.\hskip 1em plus 0.5em minus 0.4em\relax IEEE, 2021, pp. 25--2.

\bibitem{huang2021computing}
Q.~Huang \emph{et~al.}, ``Computing-in-memory using ferroelectrics: From single-to multi-input logic,'' \emph{IEEE D\&T}, vol.~39, no.~2, pp. 56--64, 2021.

\bibitem{cai2022energy}
J.~Cai \emph{et~al.}, ``Energy efficient data search design and optimization based on a compact ferroelectric fet content addressable memory,'' in \emph{ACM/IEEE DAC}, 2022, pp. 751--756.

\bibitem{yin2022ferroelectric}
X.~Yin \emph{et~al.}, ``Ferroelectric ternary content addressable memories for energy-efficient associative search,'' \emph{IEEE TCAD}, vol.~42, no.~4, pp. 1099--1112, 2022.

\bibitem{huang2023fefet}
Q.~Huang \emph{et~al.}, ``Fefet based in-memory hyperdimensional encoding design,'' \emph{IEEE TCAD}, 2023.

\bibitem{liu2022cosime}
C.-K. Liu \emph{et~al.}, ``Cosime: Fefet based associative memory for in-memory cosine similarity search,'' in \emph{IEEE/ACM ICCAD}, 2022, pp. 1--9.

\bibitem{xu2023challenges}
H.~Xu \emph{et~al.}, ``On the challenges and design mitigations of single transistor ferroelectric content addressable memory,'' \emph{IEEE EDL}, 2023.

\bibitem{yin2023ultracompact}
X.~Yin \emph{et~al.}, ``An ultracompact single-ferroelectric field-effect transistor binary and multibit associative search engine,'' \emph{Advanced Intelligent Systems}, p. 2200428, 2023.

\bibitem{yin2024ferroelectric}
X.~Yin \emph{et~al.}, ``Ferroelectric compute-in-memory annealer for combinatorial optimization problems,'' \emph{Nature Communications}, vol.~15, no.~1, p. 2419, 2024.

\bibitem{ni2018A}
K.~Ni \emph{et~al.}, ``A circuit compatible accurate compact model for ferroelectric-fets,'' in \emph{IEEE VLSI}, 2018, pp. 131--132.

\bibitem{poremba2015destiny}
M.~Poremba \emph{et~al.}, ``Destiny: A tool for modeling emerging 3d nvm and edram caches,'' in \emph{DATE}.\hskip 1em plus 0.5em minus 0.4em\relax EDA Consortium, 2015, pp. 1543--1546.

\bibitem{soliman2023first}
T.~Soliman \emph{et~al.}, ``First demonstration of in-memory computing crossbar using multi-level cell fefet,'' 2023.

\bibitem{saito2021analog}
D.~Saito \emph{et~al.}, ``Analog in-memory computing in fefet-based 1t1r array for edge ai applications,'' in \emph{2021 Symposium on VLSI Technology}.\hskip 1em plus 0.5em minus 0.4em\relax IEEE, 2021, pp. 1--2.

\bibitem{nashpyproject}
\BIBentryALTinterwordspacing
{ {The Nashpy project developers} }, ``Nashpy: <release title>.'' [Online]. Available: \url{http://dx.doi.org/10.5281/zenodo.<DOI NUMBER>}
\BIBentrySTDinterwordspacing

\end{thebibliography}

\end{document}